\documentclass[11pt,onecolumn]{IEEEtran}%
\usepackage{amsmath,amssymb,eucal,graphicx}
\usepackage{epsfig}
\usepackage{exscale}
\usepackage{latexsym}
\usepackage{verbatim}
\usepackage{amsmath}
\usepackage{amsfonts}
\usepackage{amssymb}
\usepackage{graphicx}%
\renewcommand{\QED}{\QEDopen}

\newfont{\bbb}{msbm10 scaled 500}

\newfont{\bb}{msbm10 scaled 1100}
\newcommand{\CC}{\mbox{\bb C}}
\newcommand{\RR}{\mbox{\bb R}}

\newcommand{\EE}{\mbox{\bb E}}

\newcommand{\fv}{{\bf f}}
\newcommand{\gv}{{\bf g}}
\newcommand{\hv}{{\bf h}}

\newcommand{\nv}{{\bf n}}

\newcommand{\pv}{{\bf p}}

\newcommand{\rv}{{\bf r}}
\newcommand{\sv}{{\bf s}}

\newcommand{\wv}{{\bf w}}
\newcommand{\vv}{{\bf v}}
\newcommand{\xv}{{\bf x}}
\newcommand{\yv}{{\bf y}}

\newcommand{\zerov}{{\bf 0}}

\newcommand{\Am}{{\bf A}}

\newcommand{\Gm}{{\bf G}}
\newcommand{\Hm}{{\bf H}}
\newcommand{\Id}{{\bf I}}

\newcommand{\Pm}{{\bf P}}
\newcommand{\Qm}{{\bf Q}}

\newcommand{\Sm}{{\bf S}}

\newcommand{\Cc}{{\cal C}}

\newcommand{\Nc}{{\cal N}}

\newcommand{\Pc}{{\cal P}}

\newcommand{\Sc}{{\cal S}}

\newcommand{\Vc}{{\cal V}}

\newcommand{\gammav}{\hbox{\boldmath$\gamma$}}

\newcommand{\diag}{{\hbox{diag}}}
\renewcommand{\det}{{\hbox{det}}}

\newcommand{\trace}{{\hbox{tr}}}

\renewcommand{\arg}{{\hbox{arg}}}

\renewcommand{\Re}{{\rm Re}}

\newcommand{\eqdef}{\stackrel{\Delta}{=}}

\def\BibTeX{{\rm B\kern-.05em{\sc i\kern-.025em b}\kern-.08em
T\kern-.1667em\lower.7ex\hbox{E}\kern-.125emX}}

\begin{document}

\title
{Impact of CSI on Distributed Space-Time Coding in Wireless Relay Networks}
\author{\authorblockN{Mari Kobayashi,}
\authorblockA{SUPELEC \\
Gif-sur-Yvette, France\\
Email: {\tt mari.kobayashi@supelec.fr}\\
} \and\authorblockN{Xavier Mestre,}
\authorblockA{CTTC \\
Barcelona, Spain\\
Email: {\tt xavier.mestre@cttc.cat}\\
} }
\maketitle\begin{abstract}
We consider a two-hop wireless network where a transmitter communicates with a receiver via $M$ relays with an amplify-and-forward (AF) protocol.
Recent works have shown that sophisticated linear processing such as beamforming and distributed space-time coding (DSTC) at relays enables to improve the AF performance. However, the relative utility of these strategies depend on the available channel state information at transmitter (CSIT), which in turn depends on system parameters such as the speed of the underlying fading channel and that of training and feedback procedures. Moreover, it is of practical interest to have a single transmit scheme that handles different CSIT scenarios.
This motivates us to consider a unified approach based on DSTC that potentially provides diversity gain with statistical CSIT and exploits some additional side information if available.
Under individual power constraints at the relays, we optimize the amplifier power allocation such that pairwise error probability conditioned on the available CSIT is minimized. Under perfect CSIT we propose an on-off gradient algorithm that efficiently finds a set of relays to switch on. Under partial and statistical CSIT, we propose a simple waterfilling algorithm that yields a non-trivial solution between maximum power allocation and a generalized STC that equalizes the averaged amplified noise for all relays. Moreover, we derive closed-form solutions for $M=2$ and in certain asymptotic regimes that enable an easy interpretation of the proposed algorithms. It is found that an appropriate amplifier power allocation is mandatory for DSTC to offer sufficient diversity and power gain in a general network topology.
\end{abstract}%
%EndExpansion

\section{Motivation}

%Problem setting

\begin{figure}[n]
\begin{center}
\epsfxsize=3in \epsffile{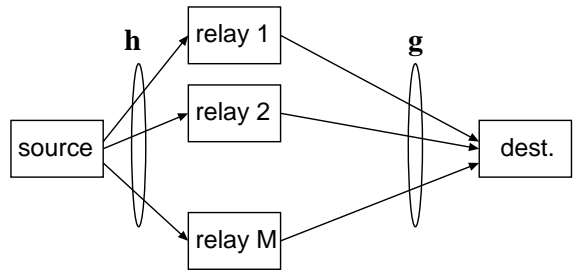}
\end{center}
\caption{A wireless relay network}%
\label{fig:model}%
\end{figure}

We consider a wireless relay network illustrated in Fig.\ref{fig:model} where
a transmitter communicates with a receiver via $M$ relays and each terminal
has a single antenna. We let $\hv,\gv$ denote the channel vector between the
transmitter and the relays, the channel vector between the relays and the
receiver.
For its simplicity, we focus on an amplify-and-forward (AF) protocol
\cite{laneman2004cdw,laneman2000eea} in a two-hop communication: in the first
$T$ channel uses the transmitter broadcasts a codeword, then in the second $T$
channel uses relays amplify and forward the observed codeword by applying some
linear precoder (to be specified later). We assume that a transmitter and $M$
relays have individual power constraints rather than a total power constraint,
%The individual power constraint is more meaningful
since in a practical wireless
network terminals are physically distributed and hence are subject to
their own power supplies.

%Existing works
In a classical AF protocol, the amplifier coefficients have been determined so
as to satisfy required power constraints %(either long-term or short-term)
\cite{laneman2004cdw}. In order to improve the performance of the AF protocol,
a large number of recent works have considered some additional linear
processing at relays \cite{WHKvtc2004,WHKglobe2004,
jing2007nbu,li2007dap,AFPrecoding, jing2007uoa, Barbarossa,DistLDC}. These
works can be roughly classified into two classes according to their assumption
of channel state information at transmitter (CSIT) and their objective. The
first class assumes perfect CSIT and aims to maximize either the achievable
rate or the instantaneous receive SNR \cite{WHKvtc2004,WHKglobe2004,
jing2007nbu,li2007dap}. The resulting transmit scheme yields beamforming with
appropriate power allocation. The second class assumes only statistical CSIT
and aims to minimize the error probability by designing some type of linear
precoder \cite{AFPrecoding, jing2007uoa, Barbarossa,DistLDC}. In particular,
significant attention has been paid to distributed space-time coding (DSTC) in
which each relay sends a different column of a STC matrix
\cite{laneman2003dst, jing2007uoa,Barbarossa,DistLDC}. %A large number of
With a single antenna at each terminal, we hasten to say that the goal of DSTC is to
approach full diversity gain of $M$ offered by the relay-receiver channels
$\gv$ since the multiplexing gain of the MISO channel under the half-duplex
constraint is very limited, i.e. 1/2. Notice that an AF based DSTC that
achieves the optimal diversity-multiplexing tradeoff \cite{zheng2003dam} has been well studied
\cite{nabar2004frc,azarian2005adm,yang2007toa,yang2007ost}.
It clearly appears that a practical utility of these two approaches depends on the available CSIT, which in turn depends on the speed of fading and that of
feedback and training procedures. Notice that in a two-hop communication model
obtaining perfect CSIT is
rather challenging because the transmitter needs at least a two-step training
and feedback process, i.e. first learns $\hv$ and then $\gv$,
which typically induces additional delay and estimation error. Consequently, the perfect CSIT assumption holds only if
the underlying fading is quasi-static and a sufficiently fast feedback and
training is available. For this particular case, the first approach might be
useful. On the contrary, if a rate of feedback and training is much slower
than the coherence time of the channel, the second approach based on the
statistical channel knowledge is more appropriate.

%Our approach
The above observation motivates us to find a unified approach that can handle
different CSIT scenarios, rather than changing a transmit strategy as a
function of the quality of side information.
To this end, we fix our transmission strategy to DSTC that potentially provides
diversity gain with statistical CSIT and further power gain if additional side
information is available. Our goal is not to find the optimal strategy for
each CSIT case but to propose a unified DSTC scheme that simply adapts the
amplifier power allocation to available CSIT.
Among a large family of DSTC, we consider linear dispersion (LD) codes
\cite{hassibi2002hrc}
because they
offer desirable performance in terms of diversity gain and coding gain
\cite{DistLDC,Jing} and moreover keep the amplified noise white. The latter
considerably simplifies the power allocation strategy. We assume perfect
synchronization between relay terminals and perfect channel state information
available at the receiver, which is necessary for coherent detection. Under
this setting, we will address the following question: how does the quality of
CSIT impact the amplifier power allocation and the resulting performance of
the DSTC? To answer the question, we optimize the amplifier power allocation
in such a manner that the pairwise error probability (PEP) conditioned on
CSIT is minimized.
Note that the conditional PEP is a performance
criteria widely used in the literature of STC
\cite{tarokh1998stc,guey1999sdt,BF-STC} and DSTC \cite{yiu2006dst,DistLDC,Jing}. In particular,
\cite{BF-STC} has provided elegant precoder designs that minimize the PEP
conditioned on CSIT for orthogonal STC. Unfortunately the extension of this work to a two-hop relay network appears very difficult due to the non-convexity of the underlying problem. We
examine the following CSIT cases : 1) perfect knowledge of the absolute value
of the entries of $\hv$ and $\gv$ (\textit{perfect CSIT}), 2) perfect
knowledge of absolute value of the entries of $\hv$ and statistical knowledge
of $\gv$ (\textit{partial CSIT}), and 3) statistical knowledge of $\hv$ and
$\gv$ (\textit{statistical CSIT}).

Under perfect CSIT, the PEP minimization reduces to the maximization of an
approximate receive SNR. The optimal power allocation strategy turns out to be an on-off strategy, whereby
some relays are switched off and others transmit at maximum available power.
We propose an on-off gradient algorithm that efficiently finds the optimal set
of relays to switch on. Under partial and statistical CSIT, the conditional
PEP minimization appears very difficult due to the self-interference caused by
amplified noise and calls for a good heuristic approximation. First, we apply a
Laplace-based saddle point approximation of an inherent integral in order to
make the problem amenable. Since
the approximated problem is still non-convex due to
{\it averaged} amplified noises, we transform it into a convex problem via a log transformation \cite{GP} assuming a high transmitter power (which is the regime
of interest). For a new objective function,
we propose a very simple waterfilling algorithm that yields a non-trivial
solution between maximum power allocation and a generalized STC that equalizes the averaged amplified noise, i.e. $p_i\gamma_{g,i}$
with $\gamma_{g,i}$ being the variance of the channel between relay $i$ and the receiver (in a classical STC we consider $\gamma_{g,i}=1,\forall i$). We derive closed-form solutions for
$M=2$ and in certain asymptotic regimes that enable an easy
interpretation of the proposed algorithms. It is found that that an appropriate
power allocation is mandatory for DSTC in order to
provide diversity and power gains in a general network topology.

%Relevance and difference compared to existing works
In order to situate this work in the context of relevant literature, we note
that the LD based DSTC for a two-hop AF network has been addressed
for a single-antenna case \cite{DistLDC} and for a
multiple antenna case \cite{Jing}. In both works, Jing and Hassibi provided
the diversity analysis by optimizing the power partition between the
transmitter and the relays under the assumptions that the transmitter
and $M$ relays are subject to a total power constraint and that both channels have unit variance. Clearly, the optimal power partition
under this setting, letting the transmitter use half of the total power and
each relay share the other half, does not hold for a general network topology
with unequal variances. We make a progress with
respect to this since our proposed waterfilling solution can be applied
for any set of variances under statistical CSIT and moreover handles the
partial CSIT case. With perfect CSIT,
Jing proposed a cooperative beamforming scheme for the same two-hop AF network
\cite{jing2007nbu}. Although this beamforming scheme provides a non-negligible power
gain compared to our on-off power allocation as shown in Section \ref{sect:Result},
we remark that our on-off algorithm is much simpler and can be implemented at the receiver without
requiring any knowledge at the transmitter. To this end, it suffices that the
receiver sends to each relay a feedback of one bit indicating whether to
activate or not. Hence, our on-off algorithm might be appealing due to its
robustness and simplicity despite its suboptimal performance.

%organization of sections
The rest of the paper is organized as follows. After briefly introducing the
two-hop network model in Section \ref{sect:Model}, we derive the the conditional PEP upper bounds for different CSIT cases in Section
\ref{sect:ConditionalPEP}. In Section \ref{sect:Algorithm} we propose
efficient algorithms that solve the conditional PEP minimization, namely
on-off gradient algorithm for perfect CSIT and waterfilling algorithm for
partial and statistical CSIT. We provide some asymptotic properties of these
algorithms in Section \ref{sect:AsymptoticBehaviors} and numerical examples in
Section \ref{sect:Result}. Finally we conclude the paper in Section
\ref{sect:Conclusions}.

%%%%%%%%%%%%%%%%%%%%%%%%%%%%%%%%%%%%%%%%%%%%%%%%%%%%%%%%%%%%%%%%%%%%%%%%%%%%%%%%%%%%%%%%%%%%%%
%%%%%%%%%%%%%%%%%%%%%%%%%%%%%%%%%%%%%%%%%%%%%%%%%%%%%%%%%%%%%%%%%%%%%%%%%%%%%%%%%%%%%%%%%%%%%%

\section{System Model}

\label{sect:Model} We consider frequency-flat fading channels and let
$\gv=[g_{1},\dots,g_{M}]^{T}$, $\hv=[h_{1},\dots,h_{M}]^{T}$ denote the
channel vector between transmitter and relays, the channel vector between
relays and receiver, respectively. We assume the entries of $\hv$ and $\gv$
are i.i.d. zero-mean circularly symmetric complex Gaussian with variance
$\gammav_{h}=[\gamma_{h1},\dots,\gamma_{hM}],\gammav_{g}=[\gamma_{g1}%
,\dots,\gamma_{gM}]$ respectively. The variance of each channel is assumed to
capture path-loss and shadowing. We assume a block fading model, namely $\hv$
and $\gv$ remain constant over a block of $2T$ channel uses. In this paper we
do not consider a transmitter-receiver direct link for simplicity. It is well
known however that the direct link should be taken into account if one aims at
optimizing the diversity-multiplexing tradeoff \cite{nabar2004frc,azarian2005adm,yang2007ost,yang2007toa}.
%NAF protocol
The communication between the transmitter and the receiver is performed in two
steps. The transmitter first broadcasts a symbol vector $\sv=[s_{1}%
,\dots,s_{T}]^{T}\in\CC^{T\times1}$ with $\EE[\sv\sv^H]=\Id_{T}$ and relay $i$ receives
\[
\yv_{i}=\sqrt{p_{s}}h_{i}\sv+\nv_{i}%
\]
where $p_{s}$ is the power of the transmitter and $\nv_{i}\sim\Nc_{\CC}%
(\zerov,N_{0}\Id_{T})$ is AWGN. In the second $T$ channel uses, $M$ relays
amplify and forward the observed codeword by applying a linear precoder.
Namely, the transmit vector $\xv_{i}$ of relay $i$ is given by
\begin{equation}
\xv_{i}=q_{i}\Am_{i}\yv_{i} \label{xi}%
\end{equation}
where $q_{i}$ denote a complex amplifier coefficient of relay $i$, $\Am_{i}%
\in\CC^{T\times T}$ is a unitary matrix satisfying $\Am_{i}^{H}\Am_{i}%
=\Id_{T}$, and $\xv_{i}$ should satisfy a power constraint, i.e.
\[
\EE[||\xv_i||^2]=|q_{i}|^{2}\EE[||\yv_i||^2]\leq Tp_{r}%
\]
where the expectation is with respect to $\nv_{i}$ for a short-term constraint
and with respect to both $\nv_{i}$ and $\hv_{i}$ for a long-term constraint.
This in turns imposes a constraint on $\{q_{i}\}$ such that $|q_{i}|^{2}\leq
P_{i}$ where $P_{i}$ denotes the maximum amplifier power of relay $i$ given
by
\[
P_{i}=\left\{
\begin{array}
[c]{ll}%
\frac{p_{r}}{p_{s}|h_{i}|^{2}+N_{0}}, & \quad\mbox{short-term }\\
\frac{p_{r}}{p_{s}\gamma_{hi}+N_{0}}, & \quad\mbox{long-term}
\end{array}
\right.
\]
The received signal at
the final destination is given by
\begin{align*}
\rv  &  =\sum_{i=1}^{M}g_{i}\xv_{i}+\wv =\sum_{i=1}^{M}g_{i}h_{i}q_{i}\Am_{i}\sv+\sum_{i=1}^{M}g_{i}q_{i}\Am_{i}\nv_{i}+\wv
\end{align*}
where $\wv\sim\Nc_{\CC}(\zerov,N_{0}\Id_{T})$ is AWGN at the receiver
uncorrelated with $\{\nv_{i}\}$. The received vector can be further simplified
to
\begin{equation}
\rv=\sqrt{p_{s}}\Sm\Qm\fv+\vv \label{r}%
\end{equation}
where we let $\Sm=[\Am_{1}\sv,\dots,\Am_{M}\sv]\in\CC^{T\times M}$ denote a LD
codeword, $\Qm=\diag(q_{1},\dots,q_{M})$ is a diagonal matrix with $M$
amplifier coefficients, $\fv=[h_{1}g_{1},\dots,h_{M}g_{M}]^{T}$ is a composite
channel vector, and we let
\begin{equation}
\vv=\sum_{i=1}^{M}q_{i}g_{i}\Am_{i}\nv_{i}+\wv
\end{equation}
denote the overall noise whose covariance is given by
\[
\EE[\vv\vv^H]=N_{0}\left(  \sum_{i=1}^{M}|q_{i}|^{2}|g_{i}|^{2}+1\right)
\Id_{T}=\sigma_{v}^{2}\Id_{T}%
\]
It follows that the overall noise seen by the receiver is white and this
considerably simplifies the amplifier power allocation in the following sections.
%Notice that the overall noise variance $\sigma_v^2$ is a function of the amplifier powers.

%%%%%%%%%%%%%%%%%%%%%%%%%%%%%%%%%%%%%%%%%%%%%%%%%%%%%%%%%%%%%%%%%%%%%%%%%%%%%%%%%%%%%%%%%%%%%%
%%%%%%%%%%%%%%%%%%%%%%%%%%%%%%%%%%%%%%%%%%%%%%%%%%%%%%%%%%%%%%%%%%%%%%%%%%%%%%%%%%%%%%%%%%%%%%
\vspace{-1em}

\section{Conditional PEP}

\label{sect:ConditionalPEP} With perfect knowledge of both $\hv$ and $\gv$,
the receiver can perform Maximum Likelihood decoding \footnote{In practice,
efficient decoding techniques such as sphere decoding can be implemented to
achieve near ML results \cite{hassibi2002hrc}.} by estimating a codeword
according to
\begin{equation}
\hat{\Sm}=\arg\min_{\Sm\in\Sc}||\rv-\sqrt{p_{s}}\Sm\Qm\fv||^{2}%
\end{equation}
When the transmitter has only partial knowledge of the channels, it is
reasonable to consider the pairwise error probability (PEP) conditioned on the available CSIT. In the following we derive the expressions of the conditional
PEP for three different CSIT cases: 1) perfect CSIT, where the transmitter
knows the absolute values of the entries of $\hv,\gv$, 2) partial CSIT where
the transmitter knows the absolute values of the entries of $\hv$ and
$\gammav_{g}$, 3) statistical CSIT where the transmitter knows $\gammav_{h}$
and $\gammav_{g}$. Perfect CSIT corresponds to the case of quasi-static
fading, while statistical CSIT corresponds to the case of fast fading so that
the transmitter can track only the second order statistics of the channel.
Finally, partial CSIT is an intermediate case relevant
to a time-division duplexing system where the transmitter learns perfectly $\hv$ by reciprocity
but only statistically $\gv$ due to a low-rate feedback.

%%%%%%%%%%%%%%%%%%%%%%%%%%%%%%%%%%%%%%%%%%%%%

\vspace{-1em}
\subsection{Perfect CSIT}

The PEP conditioned on $\hv,\gv$ for any $k\neq l$ is defined by
\begin{eqnarray*}
P(\Sm_{k}\rightarrow\Sm_{l}|\hv,\gv)&  \eqdef& \Pr\left(  ||\rv-\sqrt{p_{s}}\Sm_{l}\Qm\Hm\gv||^{2}\leq
||\rv-\sqrt{p_{s}}\Sm_{k}\Qm\Hm\gv||^{2}|\Sm_{k},\hv,\gv\right) =\Pr\left(  d^{2}(\Sm_{k},\Sm_{l})\leq\kappa\right)
\end{eqnarray*}
where the composite channel $\fv$ is decoupled into $\Hm\gv$ with
$\Hm=\diag(\hv)$, where we define squared Euclidean distance between $\Sm_{l}$
and $\Sm_{k}$ as
\[
d^{2}(\Sm_{k},\Sm_{l})=p_{s}\gv^{H}\Hm^{H}\Qm^{H}(\Sm_{k}-\Sm_{l})^{H}%
(\Sm_{k}-\Sm_{l})\Qm\Hm\gv
\]
and where $\kappa=2\sqrt{p_{s}}\Re\{\vv^{H}(\Sm_{k}-\Sm_{l})\Qm\fv\}$ is a
real Gaussian random variable distributed as $\Nc_{\RR}(0,2\sigma_{v}^{2}%
d^{2})$. In order to obtain a upper bound of the PEP, we assume that the term
$(\Sm_{k}-\Sm_{l})^{H}(\Sm_{k}-\Sm_{l})$ has a full rank $M$, i.e. the LD code
achieves a full diversity (for a special case of orthogonal STC this always holds). By letting $\lambda_{min}$ denote the smallest
singular value of $(\Sm_{k}-\Sm_{l})^{H}(\Sm_{k}-\Sm_{l})$ over all possible
codewords, we obtain the inequality
\[
d^{2}(\Sm_{k},\Sm_{l})\geq p_{s}\lambda_{min}\gv^{H}\Hm^{H}\Qm^{H}\Qm\Hm\gv
\]
%By using the above inequality, the conditional PEP is upper bounded by
which yields a Chernoff bound
\begin{equation}
P(\Sm_{k}\rightarrow\Sm_{l}|\hv,\gv)\leq\exp\left(  -\frac{p_{s}\lambda
_{min}\gv^{H}\Hm^{H}\Qm^{H}\Qm\Hm\gv}{4\sigma_{v}^{2}}\right)  .
\label{2ConditionedPEP}%
\end{equation}
Minimizing the RHS of (\ref{2ConditionedPEP}) corresponds to maximizing
\textit{approximated} receive SNR, given by
\begin{equation}
\frac{p_{s}\lambda_{\min}\gv^{H}\Hm^{H}\Qm^{H}\Qm\Hm\gv}{4\sigma_{v}^{2}}%
=\eta\frac{\sum_{i=1}^{M}p_{i}|g_{i}|^{2}|h_{i}|^{2}}{\sum_{i=1}^{M}%
p_{i}|g_{i}|^{2}+1} \label{Objective1}%
\end{equation}
where we let $\eta=\lambda_{min}p_{s}/4N_{0}$ and let $p_{i}=|q_{i}|^{2}$
denote the amplifier power of relay $i$. Notice that the above function
depends on the absolute values of channels and of amplifier coefficients.

%%%%%%%%%%%%%%%%%%%%%%%%%%%%%%%%%%%%%%%%%%%%%%%%%%%%%%%%%%%%%%%%%%%%%%%%%%%%%%

\vspace{-1em}
\subsection{Partial CSIT}
The PEP upper bound conditioned on $\hv,\gammav_{g}$ is obtained by
averaging (\ref{2ConditionedPEP}) over the distribution of $\gv$
\begin{eqnarray}
  P(\Sm_{k}\rightarrow\Sm_{l}|\hv,\gammav_{g}) %\overset{\mathrm{(a)}}{\leq}\int\frac{1}{\det(\pi\diag(\gammav_g))}\exp\left\{
%-\left(  \eta\frac{\gv^{H}\Hm^{H}\Pm\Hm\gv}{\sum_{i=1}^{M}|g_{i}|^{2}p_{i}%
%+1}+\gv^{H}\diag(\gammav_g)^{-1}\gv\right)  \right\}  d\gv\nonumber\\
&  \overset{\mathrm{(a)}}{\approx} & \int\frac{1}{\det(\pi\diag(\gammav_g))}\exp\left\{
-\gv^{H}\left(  \frac{\eta\Hm^{H}\Pm\Hm}{1+\sum_{i=1}^{M}\gamma_{gi}p_{i}
}+\diag(\gammav_g)^{-1}\right)  \gv\right\}  d\gv\nonumber\\
&=&\det\left(  \frac{\eta}{1+\sum_{i=1}^{M}\gamma_{gi}p_{i}}\diag(\gammav_g)
\Hm^{H}\Pm\Hm+\Id_{M}\right)  ^{-1}% \nonumber\\&
=\prod_{i=1}^{M}\left(  1+\eta\frac{|h_{i}|^{2}\gamma_{gi}p_{i}}
{1+\sum_{i=1}^{M}\gamma_{gi}p_{i}}\right)  ^{-1} \label{Objective2}
\end{eqnarray}
where in (a) we let $\Pm=\diag(p_{1},\dots,p_{M})$ and apply a Laplace-based saddle point
approximation that becomes accurate as the number of relays increases without
bound (see further Appendix\ \ref{sect:AppendixSaddle}). This saddle point
approximation is inspired by the approximation method suggested in
\cite{lieberman2004lam} to evaluate the expectation of quotients of quadratic
forms in Gaussian random variables. We remark that, in order to maximize the
corresponding cost function in (\ref{Objective2}), only the absolute values of
the entries of $\hv$ are needed.

\vspace{-1.2em}
\subsection{Statistical CSIT}
The PEP upper bound conditioned on $\gammav_h,\gammav_{g}$ is obtained by
averaging (\ref{2ConditionedPEP}) over the distribution of $\hv$ and $\gv$
\begin{align}
P(\Sm_{k}\rightarrow\Sm_{l}|\gammav_{h},\gammav_{g})
%&  \overset{\mathrm{(a)}}\leq\EE_{\gv}\left[  \frac{1}{\det(\pi\diag(\gammav_h))}\int_{\hv}\exp\left(
%-\hv^{H}\left(  \frac{\eta}{\sigma_{v}^{2}}\Gm^{H}\Pm\Gm+\diag(\gammav_h)^{-1}\right)
%\hv\right)  d\hv\right] \nonumber\\
&  \leq \EE_{\gv}\left[  \det^{-1}\left(  \Id_{M}+\frac{\eta}{\trace(\Gm^{H}%
\Pm\Gm)}\diag(\gammav_h)\Gm^{H}\Pm\Gm\right)  \right] \nonumber\\
&  \overset{\mathrm{(a)}}{\approx}\EE_{\gv}\left[  \prod_{j=1}^{M}\left(
1+\frac{\eta\gamma_{hj}|g_{j}|^{2}p_{j}}{1+\sum_{i=1}^{M}\gamma_{gi}p_{i}%
}\right)  ^{-1}\right] %\nonumber\\&
 \overset{\mathrm{(b)}}{=}\prod_{j=1}^{M}\frac{1}{\rho_{j}}e^{1/\rho_{j}%
}E_{1}(1/\rho_{j})\nonumber\\
&  \overset{\mathrm{(c)}}{=}\prod_{j=1}^{M}\frac{1}{\rho_{j}}\left[
-\gamma+\ln(\rho_{j})+O\left(  \frac{1}{\rho_{j}}\right)  \right] %\nonumber\\&
  \overset{\mathrm{(d)}}{\approx}\prod_{j=1}^{M}\frac{1+\sum_{i=1}^{M}%
\gamma_{gi}p_{i}}{\eta\gamma_{gj}\gamma_{hj}p_{j}}\ln\left(  \frac{\eta
\gamma_{gj}\gamma_{hj}p_{j}}{1+\sum_{i=1}^{M}\gamma_{gi}p_{i}}\right)\label{Objective3}
\end{align}
where in (a) we apply the Laplace-based saddle-point approximation of the
integral mentioned above, (b) follows by noticing that $|g_{i}|^{2}%
/\gamma_{gi}$ is an exponential random variable with unit mean and using
\cite[3.352]{gradshteyn1988tis} where we write $\rho_{j}=\frac{\eta
\gamma_{hj}\gamma_{gj}p_{j}}{1+\sum_{i=1}^{M}\gamma_{gi}p_{i}}$ and let
$E_{1}(x)=\int_{x}^{\infty}\frac{e^{-t}}{t}dt$ denote the exponential
integral. In (c) we assume that $\rho_{j}$ is large ($\eta\rightarrow\infty$)
so that
\begin{align*}
e^{1/\rho_{j}} E_{1}\left(  \frac{1}{\rho_{j}}\right)   &  =-\gamma+\ln(\rho_{j})+\sum
_{k=1}^{\infty}\frac{(-1)^{k+1}(\rho_{j})^{-k}}{kk!}%\\&
=-\gamma+\ln(\rho_{j})+O\left(  \frac{1}{\rho_{j}}\right)
\end{align*}
where $\gamma$ is the Euler constant, and finally in (d) we assume $\ln
(\rho_{j})\gg\gamma$.

%%%%%%%%%%%%%%%%%%%%%%%%%%%%%%%%%%%%%%%%%%%%%%%%%%%%%%%%%%%%%%%%%%%%%%%%%%%%%%%%%%%%%%%%%%%%%%
%%%%%%%%%%%%%%%%%%%%%%%%%%%%%%%%%%%%%%%%%%%%%%%%%%%%%%%%%%%%%%%%%%%%%%%%%%%%%%%%%%%%%%%%%%%%%%

\section{Power allocation algorithms}

\label{sect:Algorithm} This section proposes efficient power allocation
algorithms to optimize (\ref{Objective1}),
(\ref{Objective2}), and (\ref{Objective3}).

\subsection{Perfect CSIT}
Under perfect CSIT,
the optimal $\pv^{\star}$ is obtained by maximizing
\begin{eqnarray}
f_{0}(\pv)\eqdef\frac{\sum_{i=1}^{M}\alpha_{i}p_{i}
}{1+\sum_{i=1}^{M}\beta_{i}p_{i}} \label{QuasiLP}%
\end{eqnarray}
where $p_i$ is subject to the maximum amplifier power of relay $i$, i.e. $p_i\leq P_i=\frac{p_{r}%
}{p_{s}|h_{i}|^{2}+N_{0}}$ and we let $\alpha_{i}=|h_{i}|^{2}|g_{i}|^{2}$ and
$\beta_{i}=|g_{i}|^{2}$. We remark that the linear constraints form a feasible
region $\Vc$ composed by $M$ half-spaces with $2^{M}-1$ vertices. For $M=2$
the feasible region $\Vc$ is a rectangular region with 3 vertices
$(P_{1},0),(0,P_{2}),(P_{1},P_{2})$ plus the origin. Since this problem is
quasi-linear, it is possible to transform it into a linear program. By
exploiting the structure of the problem, we propose a more efficient algorithm
to find the solution. First, we start with the following proposition.

\textbf{Proposition 1} The solution to (\ref{QuasiLP}) is always found in the
one of $2^{M}-1$ vertices of the feasible region $\Vc$. Moreover, at the
solution $\pv^{\star}$, the entries of the gradient satisfy the following
inequality for $i=1,\dots,M$.
\begin{equation}
\frac{\partial f_{0}}{\partial p_{i}}(\pv^{\star})\left\{
\begin{array}
[c]{ll}%
>0, & \quad\mbox{if $p_i^{\star}=P_i$}\\
\leq0, & \quad\mbox{if $p_i^{\star}=0$}
\end{array}
\right.  \label{SufficientCondition}%
\end{equation}

\textbf{Proof } see Appendix \ref{proof1}.

For $M=2$ we have a closed form solution of the optimal power allocation as an
obvious result of Proposition 1.

\textbf{Corollary 1 } For $M=2$, we find a closed-form solution given by
\begin{equation} \label{2relayOnOff}
(p_{1},p_{2}) = \left\{
\begin{array}
[c]{ll}%
(P_1,0) , &\quad\mbox{if $|h_{1}|^{2}<\frac{p_{r}|g_{2}|^{2}|h_{2}|^{2}}{p_{s}|h_{2}|^{2}+p_{r}%
|g_{2}|^{2}+N_{0}}$}\\
 (0,P_2) , &
\quad\mbox{if $|h_{2}|^{2}<\frac{p_{r}|g_{1}|^{2}|h_{1}|^{2}}{p_{s}|h_{1}|^{2}+p_{r}%
|g_{1}|^{2}+N_{0}}$}\\
(P_{1},P_{2}) , & \quad\mbox{if $|h_{1}|^{2}>\frac{p_{r}|g_{2}|^{2}|h_{2}|^{2}}{p_{s}|h_{2}|^{2}+p_{r}%
|g_{2}|^{2}+N_{0}}$ and $|h_{2}|^{2}>\frac{p_{r}|g_{1}|^{2}|h_{1}|^{2}}{p_{s}|h_{1}|^{2}+p_{r}%
|g_{1}|^{2}+N_{0}}$}
\end{array}
\right.
\end{equation}

\textbf{Proof } see Appendix \ref{proof2}.

\begin{figure}[n]
\begin{center}
\epsfxsize=4in \epsffile{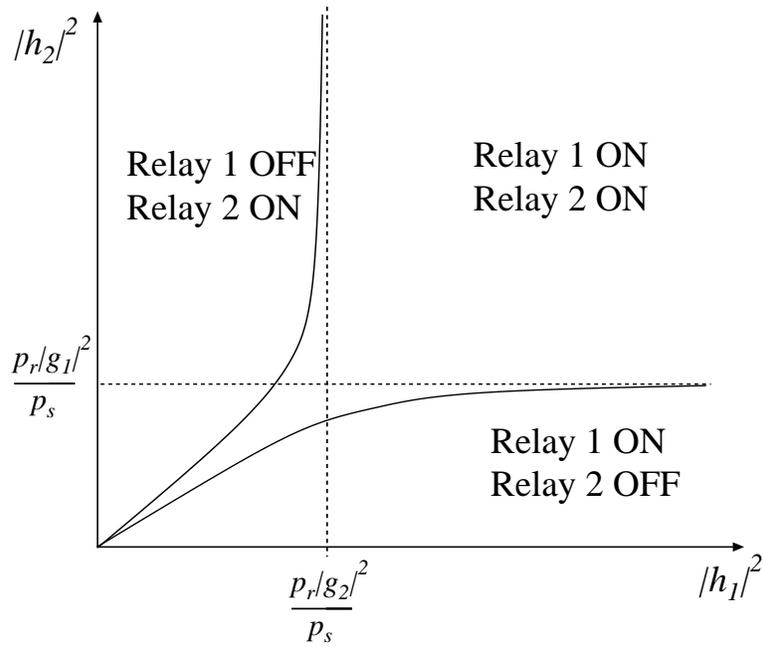}
\end{center}
\caption{Two-relay ON/OFF region under perfect CSIT}%
\label{fig:Region}%
\end{figure}

In order to visualize the conditions of
activating relay 1 and/or relay 2, we provide a graphical representation of the
on-off region in Fig. \ref{fig:Region}. Interestingly, it can be observed
there is a minimum value of $|h_{i}|^{2}$ in order for relay $i$ to be
activated. Namely relay 1 is activated independently of relay 2 if
\[
|h_{1}|^{2} > \frac{p_{r}}{p_{s}} |g_{2}|^{2}%
\]
which is readily obtained when letting $|h_{2}|^{2}\rightarrow\infty$ in the first inequality of the third condition, i.e. $|h_{1}|^{2}>\frac{p_{r}|g_{2}|^{2}|h_{2}|^{2}}{p_{s}|h_{2}|^{2}+p_{r}%
|g_{2}|^{2}+N_{0}}$, and vice versa.

For $M>2$, the solution does not lead itself to a simple closed form
expression. Nevertheless, as a straightforward result of Proposition 1 we
propose the following algorithm to solve (\ref{QuasiLP}). \newline%
\textbf{On-off gradient algorithm}

\begin{enumerate}
\item Initialize $\pv^{(0)}$ to an arbitrary vertex $\in\Vc$

\item At iteration $n$, compute the gradients $\frac{\partial f}{\partial
\pv}(\pv^{(n)})$\newline and update
\begin{equation}
\label{Update}p_{i}^{(n+1)} = \left[  p_{i}^{(n)} + \nabla_{i}(\pv^{(n)}%
)\right]  _{0}^{P_{i}},\;\;\; i=1,\dots,M
\end{equation}
where we let $\nabla_{i}(\pv^{(n)})=-P_{i}$ if $\frac{\partial f_{0}}{\partial
p_{i}}(\pv^{(n)})<0$ and $\nabla_{i}(\pv^{(n)})=P_{i}$ if $\frac{\partial
f}{\partial p_{i}}(\pv^{(n)})>0$, and
$\left[  x\right]  _{a}^{b}$ denotes the value of $x$ truncated to the
interval $[a,b]$.

\item Stop if $\pv^{(n)}$ satisfies (\ref{SufficientCondition}).
\end{enumerate}

\textbf{Proposition 2} The on-off gradient algorithm converges to the global maximum.

\textbf{Proof } See Appendix \ref{proof3}.

Fig. \ref{fig:ConvergenceM} plots the convergence behavior of the proposed
on-off algorithm when we let $p_s=p_r=10$ and consider equal variances $\gamma_{h,i}=\gamma_{g,i}=1$ for all $i$. The objective values are normalized with respect to the
optimal values and averaged over a large number of random channel
realizations.
Fig. \ref{fig:ConvergenceM} as well as other examples show that the proposed on-off
gradient algorithm converges only after a few iterations irrespectively of $M$
and of the initialization.

It is worth noticing that this on-off algorithm can be implemented at the
receiver without any knowledge at the transmitter. To this end, it suffices
that the receiver sends to each relay a feedback of one bit indicating whether
to activate or not.

%%%%%%%%%%%%%%%%%%%%%%%%%%%%%%%%%%%%%%%%%%%

\subsection{Partial CSIT}

When the transmitter knows $\hv$ and $\gammav_{g}$, the problem reduces to
minimizing (\ref{Objective2}) or maximizing
\begin{equation}
f_{1}(\pv)=\sum_{i=1}^{M}\ln\left(  1+\eta|h_{i}|^{2}\frac{\gamma_{gi}p_{i}%
}{1+\sum_{j=1}^{M}\gamma_{gj}p_{j}}\right)  \label{f1}%
\end{equation}
The term $\frac{\eta|h_{i}|^{2}\gamma_{gi}p_{i}}{1+\sum_{j=1}^{M}\gamma
_{gj}p_{j}}$, similar to $\rho_{i}=\frac{\eta\gamma_{hi}\gamma_{gi}p_{i}%
}{1+\sum_{j=1}^{M}\gamma_{gj}p_{j}}$ defined in (b) of (\ref{Objective2}) for
the statistical CSIT case, can be interpreted as the contribution of relay $i$
to the receive SNR. With some abuse of notation, we let $\rho_{i}$ denote
$\frac{\eta|h_{i}|^{2}\gamma_{gi}p_{i}}{1+\sum_{j=1}^{M}\gamma_{gj}p_{j}}$
under partial CSIT. Unfortunately, the function $f_{1}$ is neither concave or
convex in $\pv$. Nevertheless, assuming $\eta\rightarrow\infty$ (which
is the regime of our interest), let us consider a new objective function,
given by
\begin{equation}
J(\pv)=\sum_{i=1}^{M}\ln\left(  \frac{a_{i}p_{i}}{1+\sum_{j}\gamma_{gj}p_{j}%
}\right)  \label{NewObjective}%
\end{equation}
where we let $a_{i}=\eta\gamma_{gi}|h_{i}|^{2}$ for notation simplicity. It is
well known that the function $J(\pv)$ can be transformed into a
\textit{concave} function through a log transformation \cite{GP}. In the
following, we use the notation $\widetilde{x}$ to express $\ln x$ for any
variable $x$ (equivalently $x=e^{\tilde{x}}$). The objective function can be
expressed in terms of $\widetilde{\pv}$ as
\begin{equation}
J(\widetilde{\pv})=\sum_{i=1}^{M}\ln(a_{i}e^{\widetilde{p_{i}}})-M\ln\left(
1+\sum_{j}\exp(\widetilde{p_{j}})\gamma_{gj}\right)  \label{LogObjective}%
\end{equation}

\textbf{Proposition 3 } The optimal $\widetilde{\pv}$ that maximizes
(\ref{LogObjective}) is given
\begin{equation}\label{WFSolution}
\tilde{p}_{i}=\left[  \tilde{\mu}^{\star}-\tilde{\gamma}_{gi}\right]  _{-\infty
}^{\tilde{P}_{i}},\;\;\;\;p_{i}=\left[  \frac{\mu^{\star}}{\gamma_{gi}}\right]
_{0}^{P_{i}}%
\end{equation}
where $\tilde{\mu}^{\star}$ is the \textit{water level} that is
determined as follows. Let $\pi$ denote a permutation such that
\begin{equation}\label{Permutation}
P_{\pi(1)}\gamma_{g,\pi(1)}\leq\ldots\leq P_{\pi(M)}\gamma_{g,\pi(M)}.
\end{equation}
and define $\tilde{\mu_{j}}=\ln\mu_{j}$, where $\mu_{j}$ is given by
\begin{equation}\label{MuCandidate}
\mu_{j}=\frac{1+\sum_{i=1}^{j}P_{\pi(i)}\gamma_{g,\pi(i)}}{j}\;\;\;j=1,\dots
,M.
\end{equation}
The optimal water level $\tilde{\mu}^{\star}$ is obtained as that the value out of these
M possible ones that maximizes the objective function, namely
\begin{equation}\label{OptimalLevel}
\tilde{\mu}^{\star}=\arg\max_{\tilde{\mu}_{1},\dots,\tilde{\mu}_{M}}J(\tilde{\mu}_{j})
\end{equation}
where $J(\tilde{\mu})$ is the objective function (\ref{LogObjective}) parameterized by the water level defined in (\ref{Jmu}).

{\bf Proof }  see Appendix \ref{proof4}.

Fig. \ref{fig:KKT} illustrates an example of our waterfilling solution for the
case $M=3$. The power curve of relay $i$ increases linearly with slope
$1/\gamma_{gi}$ and then is bounded at its maximum amplifier power $P_{i}$. In
this example, relays 1 and 2 with $P_{i}\gamma_{gi}<\mu^{\star}$ are allocated their
maximum amplifier powers while relay $3$ is allocated $\frac{\mu^{\star}}{\gamma
_{g,3}}$. Depending on the water level, this waterfilling yields
a non-trivial solution between maximum power allocation ($p_{i}=P_{i},\forall
i$) and a generalized STC that equalizes the averaged amplified noise $p_{i}\gamma_{gi}=\mu^{\star}$ (notice a
classical STC considers $\gamma_{gi}=1$ for all $i$). Note that the proposed
waterfilling approach only requires a search over $M$ values in order to
determine the water level, and consequently it is extremely simple.

\textbf{Corollary 2 } For $M=2$, we find a closed-form solution given by
\begin{equation}
\label{3Subregion}(p_{1},p_{2}) = \left\{
\begin{array}
[c]{ll}%
\left(  \frac{1+P_{2}\gamma_{g2}}{\gamma_{g1}}, P_{2}\right)  , &
\quad\mbox{if $P_2 \gamma_{g2} \geq P_1 \gamma_{g1} +1 $}\\
\left(  P_{1}, \frac{1+P_{1}\gamma_{g1}}{\gamma_{g2}}\right)  , &
\quad\mbox{if $P_2 \gamma_{g2} \leq P_1 \gamma_{g1} -1$}\\
(P_{1},P_{2}) , & \quad\mbox{otherwise}
\end{array}
\right.
\end{equation}

\textbf{Proof } see Appendix \ref{proof5}.

By expressing the power constraint $P_{i}=\frac{p_{r}}{p_{s}|h_{i}|^{2}+N_{0}%
}$, the above allocation policy can be graphically represented as a function
of $|h_{1}|^{2}$ and $|h_{2}|^{2}$ in Fig. \ref{fig:HRegion}. Similarly to
Fig. \ref{fig:Region} for the case of perfect CSIT, there exists a minimum value of $|h_{i}|^{2}$ so that
relay $i$ is allocated its maximum amplifier power. Namely, relay $i$ is
allocated its maximum amplifier power independently of relay $j\neq i$ if
$|h_{i}|^2 >\frac{p_{r}\gamma_{g,i}-N_{0}}{p_{s}}$.
Interestingly, the threshold associated to relay $i$ depends only on the
$i$-th channel, as opposed of what happened in the perfect CSIT case. This
means that the power allocation is more selfish under partial/statistical CSIT
in order to increase the reliability of the wireless link.

%%%%%%%%%%%%%%%%%%%%%%%%%%%%%%%%%%%%%%%%%%%%%%%%%%%%%%%%%%%%%%%%%%%%%%%%%%%%%%%%%%%%%%%%%%%%%%
%%%%%%%%%%%%%%%%%%%%%%%%%%%%%%%%%%%%%%%%%%%%%%%%%%%%%%%%%%%%%%%%%%%%%%%%%%%%%%%%%%%%%%%%%%%%%%

\vspace{-2.1em}
\subsection{Statistical CSIT}

When the transmitter only knows the variances $\gammav_{g},\gammav_{h}$ of the
channels, we minimize the expression (\ref{Objective3}) which is equivalent to
maximizing
\begin{equation}
f_{2}(\pv)=\sum_{i=1}^{M}\ln\left(  \frac{\eta\gamma_{hi}\gamma_{gi}p_{i}%
}{1+\sum_{j=1}^{M}p_{j}\gamma_{gj}}\right)  \label{f2}%
\end{equation}
where we ignored doubly logarithmic terms and the amplifier power $p_{i}$ of
relay $i$ is subject to a long-term individual power constraint $P_{i}%
=\frac{p_{r}}{p_{s}\gamma_{hi}+N_{0}}$ for all $i$. Again, by performing a
log-transformation we obtain precisely the same objective function
$J(\tilde{\pv})$ in (\ref{LogObjective}) where $a_{i}=\eta\gamma_{gi}%
|h_{i}|^{2}$ defined in the previous partial CSIT case is replaced with
$\eta\gamma_{gi}\gamma_{hi}$. Hence, the waterfilling solution proposed for
the partial CSIT case can be directly applied to the statistical CSIT case and
needs to be implemented once for a given set of variances $\gammav_{h}%
,\gammav_{g}$. The power allocation region for $M=2$ is given in Fig.
\ref{fig:HRegion} where the axes are replaced by $\gamma_{h1}$ and
$\gamma_{h2}$.

%%%%%%%%%%%%%%%%%%%%%%%%%%%%%%%%%%%%%%%%%%%%%%%%%%%%%%%%%%%%%%%%%%%%%%%%%%%%%%%%%%%%%%%%%%%%%%

\section{Asymptotic behavior}\label{sect:AsymptoticBehaviors}
This section studies the asymptotic behavior
of the proposed power allocation algorithms and gives an informal discussion
on the resulting error rate performance.

\subsection{Relays close to transmitter $\gammav_{h}\rightarrow\infty$}

\label{subsect:h} We consider the regime where $\gamma_{h,i}\rightarrow\infty$
or equivalently $|h_{i}|^{2}\rightarrow\infty$ for all $i$ at the same rate
while treating other parameters finite. First we examine a two relay case.
Under prefect CSIT, Fig. \ref{fig:Region} implies that both relays tend to be
switched on in this regime. Under partial and statistical CSIT cases, it
follows immediately from Fig. \ref{fig:HRegion} that two relays shall transmit
at their maximum powers.

For $M>2$ the same conclusion can be drawn as a straightforward result of
Proposition 1 for the perfect CSIT case. The condition for which all $M$
relays are switched on under perfect CSIT is given by
\[
|h_{i}|^{2}>\frac{\sum_{j=1}^{M}P_{j}|h_{j}g_{j}|^{2}}{1+\sum_{j=1}^{M}%
P_{j}|g_{j}|^{2}},\;\;i=1,\dots,M
\]
where the RHS corresponds to the objective value $f_{0}(P_{1},\dots,P_{M})$
when letting all relays transmit with maximum power. The RHS is upper bounded
by
\begin{align}
f_{0}(P_{1},\dots,P_{M})  &  \overset{\mathrm{(a)}}{\leq}\frac{p_{r}}{p_{s}%
}\frac{\sum_{j=1}^{M}|g_{j}|^{2}}{1+\sum_{j=1}^{M}P_{j}|g_{j}|^{2}%
} \overset{\mathrm{(b)}}{\leq}\frac{p_{r}}{p_{s}}\sum_{j=1}^{M}|g_{j}|^{2}\label{UpBound}
\end{align}
where (a) follows from $P_{j}|h_{j}|^{2}=\frac{p_{r}|h_{j}|^{2}}{p_{s}%
|h_{j}|^{2}+N_{0}}\leq\frac{p_{r}}{p_{s}}$ and (b) follows from $\frac
{1}{1+\sum_{j=1}^{M}|g_{j}|^{2}P_{j}}\leq1$. In the limit of $|h_{j}%
|^{2}\rightarrow\infty,\forall j$, both (a) and (b) hold with equality and we
have
\[
|h_{i}|^{2}>\frac{p_{r}}{p_{s}}\sum_{j=1}^{M}|g_{j}|^{2},\;\;i=1,\dots,M
\]
so that all relays are allocated their maximum powers. From the upper bound
(\ref{UpBound}) of the objective function, it can be expected that the
performance of distributed LD code improves for a larger $M$. Under partial
and statistical CSIT, the proposed waterfilling tends to allocate the maximum
power to each relay. This can be seen immediately from the waterfilling
solution depicted in Fig.\ref{fig:KKT}. As $P_{j}\rightarrow0$, the values
$\{P_{i}\gamma_{g,i}\}$ above which the power curves are bounded become much
smaller than the lowest water level $\mu_{\mathrm{min}}=1/M$. This means that
all relays are allocated the maximum powers. The following remarks are in order;

\begin{enumerate}
\item Since the waterfilling algorithm under partial CSIT and on-off gradient
algorithm coincide, both algorithms yield the same error performance. This
implies that the knowledge of $\gv$ has a negligible effect on the performance
in the regime.

\item As a final remark, the same behavior can be
observed in the following cases.

\begin{itemize}
\item the transmitter power increases $p_{s}\rightarrow \infty$.

\item the variance $\gammav_{g}$ of the relay-receiver channel decreases, i.e.  $\gammav_{g}\rightarrow\zerov$ or equivalently $|g_i|^2\rightarrow 0$ for all $i$ at the same rate.
\end{itemize}

\end{enumerate}

%%%%%%%%%%%%%%%%%%%%%%%%%%%%%%%%%%%%%%%%%%%%%%%%%%%%%%%%%%%%%%%%%%%%%%%%%%

\subsection{Relays get close to receiver $\gammav_{g}\rightarrow\infty$}

\label{subsect:g} We consider the regime where $\gamma_{g,i}\rightarrow\infty$
or equivalently $|g_{i}|^{2}\rightarrow\infty$ for all $i$ at the same rate.
First we examine a two-relay case $M=2$. Under perfect CSIT, it can be
observed that the threshold values $\frac{p_{r}|g_{i}|^{2}}{p_{s}}$ above
which each relay becomes activated (represented by straight lines in Fig.
\ref{fig:Region}) get large and the on-off algorithm converges to relay selection.
Similarly, the waterfilling algorithm also tends to allocate only one relay
with maximum power under partial and statistical CSIT cases as expected from
Fig. \ref{fig:HRegion}. The only exception is a symmetric variance case
$\gamma_{h,1}=\gamma_{h,2}$.

For $M>2$ under perfect CSIT, we next show that the on-off strategy
converges to single relay selection as $|g_{i}|^{2}\rightarrow\infty,\forall i$. To see this, first observe that if the one-off algorithm
chooses only one relay to switch on, it selects the relay
\[
i^{\star}=\arg\max_{i}\frac{|h_{i}|^{2}}{1+1/|g_{i}|^{2}P_{i}}%
\]
The objective value is given by (with some abuse of notation in the argument
of the cost function)
\begin{align}
f_{0}(P_{i^{\star}},\zerov_{M-1})  &  =\frac{|h_{i^{\star}}|^{2}%
}{1+1/|g_{i^{\star}}|^{2}P_{i^{\star}}} \leq|h_{i^{\star}}|^{2} \label{fselect}
\end{align}
where the inequality holds with equality for $|g_{i^{\star}}|^{2}%
\rightarrow\infty$. If the on-off algorithm activates any arbitrary set of $m>1$
relays, the corresponding objective is upper bounded by
\begin{align}
f_{0}(P_{1},\dots,P_{m},\zerov_{M-m})  &  =\frac{\sum_{i=1}^{m}|h_{i}%
|^{2}|g_{i}|^{2}P_{i}}{1+\sum_{j=1}^{m}|g_{j}|^{2}P_{j}} <\frac{\sum_{i=1}^{m}|h_{i}|^{2}|g_{i}|^{2}P_{i}}{\sum_{j=1}^{m}|g_{j}%
|^{2}P_{j}} \overset{\mathrm{(a)}}{=}\sum_{i=1}^{m}\theta_{i}|h_{i}|^{2}\label{mrelay}
\end{align}
where in (a) we define $\theta_{i}=\frac{|g_{i}|^{2}P_{i}}{\sum_{j=1}%
^{m}|g_{j}|^{2}P_{j}}$ with $0<\theta_{i}<1$ and $\sum_{i}\theta_{i}=1$. We
see that the last expression (\ref{mrelay}) is strictly smaller than
(\ref{fselect}) for any $m$ and regardless of a set of relays. This implies
that as the relays get close to the receiver, the on-off algorithm converges to
single relay selection.

The waterfilling algorithm under partial and statistical CSIT lets only one
relay transmit with the maximum power as we see in the following. Let us first
consider the permutation $\pi$  given in (\ref{Permutation}) sorting relays according to
$P_{\pi(1)}\gamma_{g,\pi(1)}<P_{\pi(2)}\gamma_{g,\pi(2)}<\dots<P_{\pi(M)}%
\gamma_{g,\pi(M)}$ with strict inequalities. As $\gamma_{g,i}\rightarrow\infty$ for all $i$, the
possible water level in (\ref{MuCandidate}) is roughly given by $\mu
_{j}\approx\frac{\sum_{i=1}^{j}P_{\pi(i)}\gamma_{g,\pi(i)}}{j}$ and the levels
tend to be sorted as
\begin{equation}
\mu_{\min}\ll\mu_{1}<\mu_{2}\leq\dots<\mu_{M}=\mu_{\max} \label{Order}%
\end{equation}
Notice that the inequality $\mu_{j}<\mu_{j+1}$ holds if $\sum_{i=1}^{j}%
(P_{\pi(j+1)}\gamma_{g,\pi(j+1)}-P_{\pi(i)}\gamma_{g,\pi(i)})>1$. We show that the
function $J$ is monotonically decreasing for $\mu_{1}\leq\mu\leq\mu_{M}$ and
the optimal water level is always given by $\mu_{1}$. We recall that the
derivative of $J$ with respect to $\tilde{\mu}$ in (\ref{DerivativeMu}) can be
expressed as a function of $\mu$ as
\[
\nabla J=1-\frac{M\mu}{1+|\Cc(\mu)|\mu+\sum_{i\in\overline{\Cc}(\mu)}%
\gamma_{gi}P_{i}}%
\]
where we assumed $|\Cc(\mu)|>0$. Under the specific order of the water levels
given in (\ref{Order}), we can further express the derivative $\nabla J_{j}$
for each interval $\mu_{j}<\mu\leq\mu_{j+1}$ for $j=1,\dots,M-1$ such that
\begin{align}
\nabla J_{j}  %&  =\frac{-j\mu+1+\sum_{i=1}^{j}\gamma_{g,\pi(i)}P_{\pi(i)}%
%}{1+(M-j)\mu+\sum_{i=1}^{j}\gamma_{g,\pi(i)}P_{\pi(i)}}\\
&  =\frac{j(\mu_{j}-\mu)}{1+(M-j)\mu+\sum_{i=1}^{j}\gamma_{g,\pi(i)}P_{\pi
(i)}}\nonumber
\end{align}
Since we have $\nabla J_{j}<0$ for any interval $j$, it clearly appears that
the function is monotonically decreasing thus the waterfilling algorithm
allocates maximum power only to the relay $\pi(1)$ by letting $\mu^{\star}=\mu_1$. In order to have an
insight on the error rate performance achieved by the waterfilling letting
$p_{i}=\frac{\mu_{1}}{\gamma_{gi}}$ for all $i$, we evaluate the approximated
receive SNR value $f_{0}$.
\begin{align}
f_{0}\left(  \left\{  \frac{\mu_{1}}{\gamma_{gi}}\right\}  \right)   &
=\frac{\mu_{1}\sum_{i=1}^{M}|h_{i}|^{2}\frac{|g_{i}|^{2}}{\gamma_{gi}}}%
{1+\mu_{1}\sum_{j=1}^{M}\frac{|g_{j}|^{2}}{\gamma_{gj}}}
\overset{\mathrm{(a)}}{\leq}\frac{\sum_{i=1}^{M}|h_{i}|^{2}|
g_{i}|^{2}/\gamma_{gi}}{\sum_{j=1}^{M}|g_{j}|^{2}/\gamma_{gj}} \overset{\mathrm{(b)}}{=}\sum_{i=1}^{M}\theta_{i}|h_{i}|^{2}%\label{imperfect}
\end{align}
where (a) holds with equality as $\mu_{1}\rightarrow\infty$
($P_{i}\gamma_{gi}\rightarrow\infty$ for all $i$), and in (c) we define
$\theta_{i}=\frac{|g_{i}|^2/\gamma_{gi}}{\sum_{j=1}^M|g_{j}|^{2}/\gamma_{gj}}$ with $0<\theta_{i}<1$ and $\sum_{i}\theta_{i}=1$.
We see that the final expression is dominated by (\ref{fselect})
achieved by single relay selection under perfect CSIT. The following remarks
are in order;

\begin{enumerate}
\item The optimal transmit scheme in this regime is single relay selection
that chooses roughly the relay with the largest $|h_{i}|^{2}$. Activating more
than one relay becomes highly suboptimal due to large amplified noise.

\item The power allocation under partial and statistical CSIT is the
same and equalizes $p_{i}\gamma_{gi}$.

\item As a final remark, the same behavior can be observed in the following
cases.

\begin{itemize}
\item the relay power increases $p_{r}\rightarrow\infty$.

\item the variance of the transmitter-relay channel decreases $\gammav_{h}\rightarrow \zerov$.
\end{itemize}

which yield $P_{i}|g_{i}|^{2}\rightarrow\infty$ under perfect CSIT and
$P_{i}\gamma_{g,i}\rightarrow\infty$ under partial and statistical CSIT.
\end{enumerate}

%%%%%%%%%%%%%%%%%%%%%%%%%%%%%%%%%%%%%%%%%%%%%%%%%%%%%%%%%%%%%%%%%%%%%%%%%%%%%%%%%%%%%%%%%%%%%%

\section{Numerical results}

\label{sect:Result} In this section, we provide some numerical results to
illustrate the behavior of the proposed power allocation algorithms. Assuming
a homogeneous network, we let $p_{s}=p_{r}$. We consider BPSK modulation and
generate randomly a LD code with $M=T$ drawn from an isotropic distribution.

\begin{figure}[n]
\begin{center}
\epsfxsize=4in \epsffile{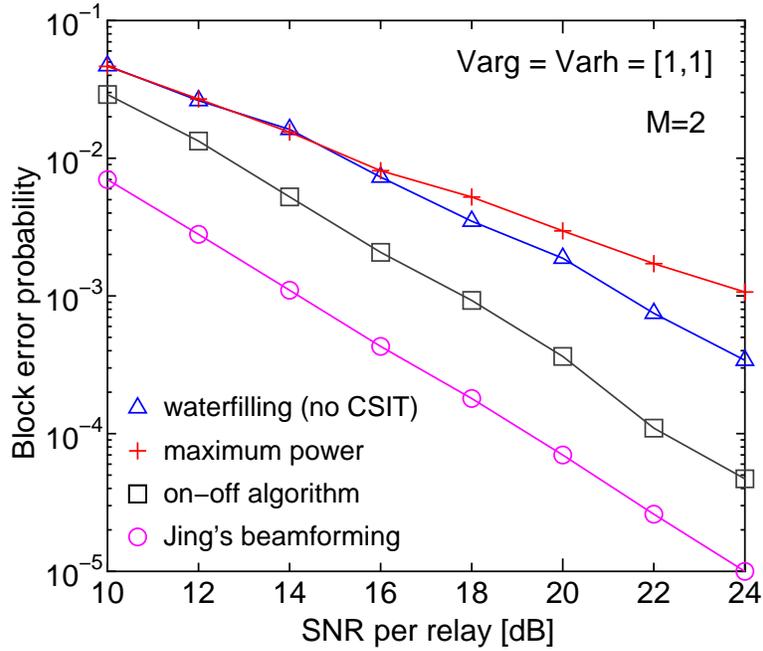}
\end{center}
\caption{Block error rate vs SNR }%
\label{fig:OnoffvsBeamformingM2}%
\end{figure}

First, we compare the proposed on-off algorithm with other schemes in a system
with $M=2$ relays and equal variances $\gamma_{hi}=\gamma_{gi}=1$ for $i=1,2$.
Fig. \ref{fig:OnoffvsBeamformingM2} shows the block error probability versus
per-relay SNR $p_{r}/N_{0}$ with the on-off algorithm, network beamforming of
\cite{jing2007nbu}, and maximum power allocation that lets both relays
transmit with their peak powers. For a reference we also plot the performance
of our waterfilling algorithm under statistical CSIT. We observe that network
beamforming outperforms the on-off gradient algorithm by roughly 3 dB by
exploiting full channel knowledge and that both schemes achieve the same
diversity gain. On the contrary, maximum power allocation has a substantial
performance loss and fails to achieve full diversity gain. This clearly shows
that an appropriate power allocation is essential for distributed LD code to
provide diversity gain.

\begin{figure}[n]
\begin{center}
\epsfxsize=4in \epsffile{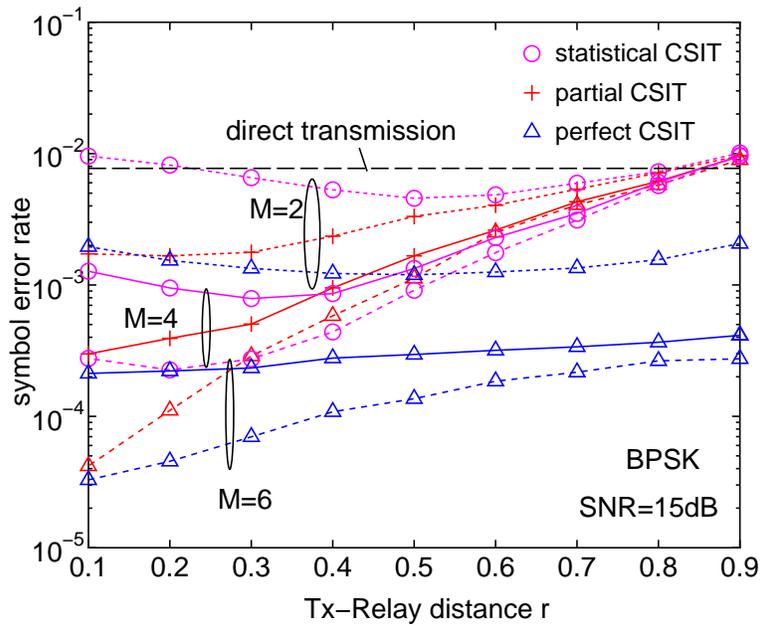}
\end{center}
\caption{BER vs. transmitter-relay distance }%
\label{fig:BERvsDistance}%
\end{figure}

\begin{figure}[n]
\begin{center}
\epsfxsize=4in \epsffile{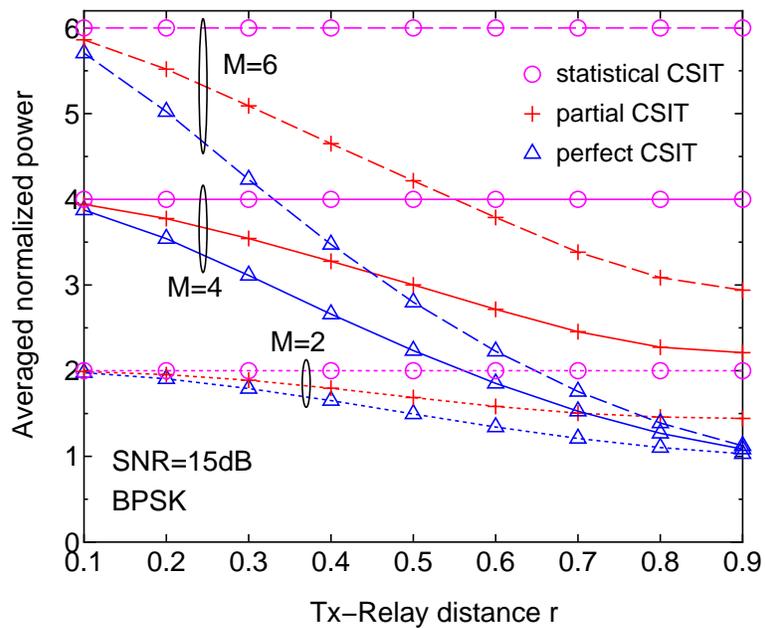}
\end{center}
\caption{Normalized allocated power vs. transmitter-relay distance }%
\label{fig:NumRelayvsR}%
\end{figure}

Next, we examine how the network topology impacts the proposed power
allocation algorithms and the resulting BER performance. To model a simple
network topology, we consider a unit transmitter-receiver distance and let the
transmitter-relay distance varies in the range $0< r <1$. The resulting
variances are $\gamma_{hi}=1/r^{2}$ and $\gamma_{gi}=1/(1-r)^{2}$ for all $i$.
For the sake of fair comparison between systems with different $M$, we assume
that the whole network power $\Pc$ is equally shared between the transmitter
and $M$ relays so that $p_{r}=\Pc/(M+1)$. Fig. \ref{fig:BERvsDistance} shows
the BER performance of the proposed power allocation algorithms with $M=2,4,6$
and $\Pc/N_{0}=15$ dB along with the performance of the direct transmission
with a fixed power $\Pc$. Fig. \ref{fig:NumRelayvsR} shows the averaged
allocated power ratio, i.e. $\sum_{i=1}^{M}\EE[p_i(t)/P_i(t)]$, or
equivalently the effective number of relays. The following remarks are in
order: 1) As the relays get closer to the transmitter $r\rightarrow0$, the
transmitter activates all relays with their maximum power. The waterfilling
solution under partial CSIT converges to the on-off gradient algorithm in the
limit of $r\rightarrow0$, which implies that the knowledge of $\gv$ has a
negligible impact on the performance. The result agrees well with the analysis
provided in subsection \ref{subsect:h}.
2) As the relays get closer to the receiver $r\rightarrow1$, the optimal
strategy activates only one relay to limit the amplified noise. As seen in
Fig. \ref{fig:NumRelayvsR} the on-off gradient algorithm indeed reduces to
relay selection. On the contrary, the waterfilling solution equalizes
$p_{1}\gamma_{g1}=\dots=p_{M}\gamma_{gM}$ both under partial and statistical
CSIT, and moreover it converges to the same error performance independently
the number of relays. Under the given setting where $P_{1}\gamma_{g1}%
=\dots=P_{M}\gamma_{gM}$, the waterfilling solution under statistical CSIT
lets all relays transmit with maximum power. The result is in a good agreement
with the analysis of subsection \ref{subsect:g}.

\begin{figure}[n]
\begin{center}
\epsfxsize=4in \epsffile{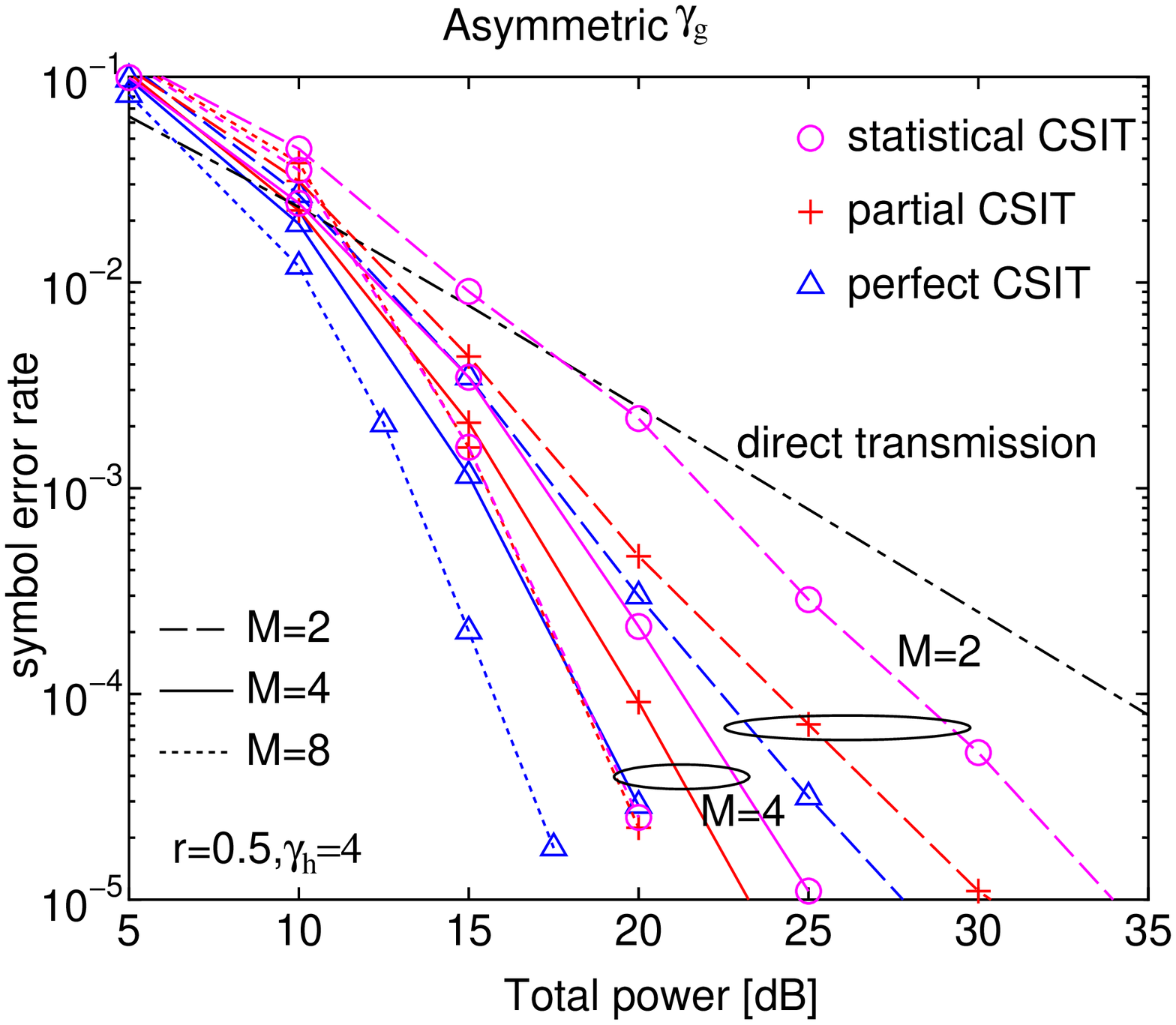}
\end{center}
\caption{BER performance vs $\Pc/N_{0}$ }%
\label{fig:BER2}%
\end{figure}

Fig. \ref{fig:BER2} shows the BER performance versus $\Pc/N_{0}$ for
$M=2,4,8$. Here, we randomly choose the relay-receiver distances and let
$\gammav_{g}$=[0.85, 3.17, 1.50, 1.89, 2.06, 2.36, 3.19, 3.99]. The
transmitter-relay distance $r=0.5$ is fixed ($\gamma_{hi}=4$ for any $i$).
Compared to the direct transmission, DSTC with our proposed power allocation
algorithms yields significant diversity gain at moderate to high power regime.
Moreover, additional CSIT yields a considerable power gain.

%%%%%%%%%%%%%%%%%%%%%%%%%%%%%%%%%%%%%%%%%%%%%%%%%%%%%%%%%%%%%%%%%%%%

\section{Conclusions}

\label{sect:Conclusions} We considered a two-hop wireless network where $M$
relays aid one transmitter-receiver pair to communicate via DSTC together with the AF protocol. In order to study the
impact of CSIT on the design and the performance of DSTC, we optimized
the amplifier power allocation under individual power constraints so that the
PEP conditioned to the available CSIT is minimized. Under
perfect CSIT we proposed the on-off gradient algorithm that efficiently finds a
subset of relays to switch on. It turns out that this algorithm can be
implemented at the receiver if the receiver can send a one-bit feedback to
each relay indicating whether to switch on or not. Under partial and
statistical CSIT we derived a simple waterfilling algorithm that yields a
non-trivial solution between maximum power allocation and a generalized STC
that equalizes the averaged amplified powers for all relays. Closed-form solutions were derived for $M=2$ and
in certain asymptotic regimes. Namely, when the relays are physically close to
the transmitter, the on-off algorithm and the waterfilling algorithm coincide and
both let all relays transmit with maximum amplifier powers. When the relay are
close to the receiver, the on-off algorithm converges to relay selection in order
to minimize the amplified noise seen by the receiver while the waterfilling
equalizes the averaged amplified noise and becomes highly suboptimal. The proposed
amplifier power allocation algorithms were derived for a particular type of
linear dispersion STC but can be extended to more general LD code as well as
other classes of STC as long as the amplified noise remains white.

\appendices

%%%%%%%%%%%%%%%%%%%%%%%%%%%%%%%%%%%%%%%%%%%%%%%%%%%%%%%%%%%%%%%%%%%%%%%%%%%%%%%%%%%%%%%%%%%%

\vspace{-0.8em}
\section{Saddle point approximation}

%%%%%%%%%%%%%%%%%%%%%%%%%%%%%%%%%%%%%%%%%%%%%%%%%%%%%%%%%%%%%%%%%%%%%%%%%%%%%%%%%%%%%%%%%%%%
\label{sect:AppendixSaddle}The objective of this appendix is to justify the
approximations (a) in (\ref{Objective2}) and
(\ref{Objective3}) and to show that the approximation is valid as the number
of relays increases without bound. Let us denote
\[
a=\mathbf{g}^{H}\mathbf{H}^{H}\mathbf{PHg\quad}b=1+\mathbf{g}^{H}\mathbf{Pg}%
\]
We need to evaluate
\[
\mathbb{E}\left[  \exp\left(  -\eta\frac{a}{b}\right)  \right]  =\mathbb{E}%
\left[  \exp\left(  -\eta\frac{\mathbf{g}^{H}\mathbf{H}^{H}\mathbf{PHg}%
}{1+\mathbf{g}^{H}\mathbf{Pg}}\right)  \right]
\]
where the expectation is with respect to the statistics of
$\mathbf{g}$. First of all, observe that we can write
\[
\exp\left(  -\eta\frac{\mathbf{g}^{H}\mathbf{H}^{H}\mathbf{PHg}}%
{1+\mathbf{g}^{H}\mathbf{Pg}}\right)  =\lim_{n\rightarrow\infty}X_{n}%
\]
where
\[
X_{n}=\sum_{k\geq0}^{n}\frac{\left(  -\eta\right)  ^{k}}{k!}\left(  \frac
{a}{b}\right)  ^{k}.
\]
On the other hand,
\begin{multline*}
\left\vert X_{n}\right\vert <\sum_{k\geq0}^{n}\frac{\eta^{k}}{k!}\left(
\frac{a}{b}\right)  ^{k}<\exp\left[  \eta\frac{\mathbf{g}^{H}\mathbf{H}%
^{H}\mathbf{PHg}}{1+\mathbf{g}^{H}\mathbf{Pg}}\right] <\exp\left[  \eta\frac{\mathbf{g}^{H}\mathbf{H}^{H}\mathbf{PHg}}%
{\mathbf{g}^{H}\mathbf{Pg}}\right]  \leq\exp\left(  \eta\max_{1\leq j\leq
M} \left\vert h_{j}\right\vert ^{2} \right)  <\infty
\end{multline*}
Hence, the bounded convergence theorem ensures that we can write
\begin{equation}
\mathbb{E}\left[  \exp\left(  -\eta\frac{a}{b}\right)  \right]  =\mathbb{E}%
\left[  \lim_{n\rightarrow\infty}X_{n}\right]  =\lim_{n\rightarrow\infty
}\mathbb{E}\left[  X_{n}\right]  =\sum_{k\geq0}\frac{\left(  -\eta\right)
^{k}}{k!}\mathbb{E}\left[  \left(  \frac{a}{b}\right)  ^{k}\right]
\label{bounded_convergence}%
\end{equation}
and we can therefore concentrate on the study of the moments
\[
r_{k}=\mathbb{E}\left[  \left(  \frac{a}{b}\right)  ^{k}\right]  .
\]
In particular, we can follow the procedure introduced in \cite{lieberman2004lam},
which is based on a Laplace approximation of the integral. More specifically, in \cite{lieberman2004lam} it was shown that a Laplace approximation of $r_{k}$ about the origin leads to the identity%
\begin{equation}
r_{k}=\mathbb{E}\left[  \left(  \frac{a}{\mathbb{E}\left[  b\right]  }\right)
^{k}\right]  +R_{k}\label{rk_asym}%
\end{equation}
where both expectations are with respect to $\gv$ and we have
$R_{k}\rightarrow0$ as $M\rightarrow\infty$. Although the procedure is
somewhat tedious, one can extend this to show that $\sup_{k}R_{k}\rightarrow
0$. As a direct consequence of this theorem, we have that as $M\rightarrow
\infty$ , it holds that\footnote{In the last equality of the following
section, one should justify the fact that expectation and sum can be
interchanged. This is not difficult to see, but the proof is omitted due to
space constraints.}%
\begin{multline*}
\mathbb{E}\left[  \exp\left(  -\eta\frac{a}{b}\right)  \right]  =\sum_{k\geq
0}\frac{\left(  -\eta\right)  ^{k}}{k!}\mathbb{E}\left[  \left(  \frac{a}%
{b}\right)  ^{k}\right]  \\
=\sum_{k\geq0}\frac{\left(  -\eta\right)  ^{k}}{k!}\left(  \mathbb{E}\left[
\left(  \frac{a}{\mathbb{E}\left[  b\right]  }\right)  ^{k}\right]
+R_{k}\right)  =\mathbb{E}\left[  \exp\left(  -\eta\frac{a}{\mathbb{E}\left[
b\right]  }\right)  \right]  +\sum_{k\geq0}\frac{\left(  -\eta\right)  ^{k}%
}{k!}R_{k}%
\end{multline*}
where now
\[
\left\vert \sum_{k\geq0}\frac{\left(  -\eta\right)  ^{k}}{k!}R_{k}\right\vert
\leq\sup_{k}R_{k}\exp\left(  -\eta\right)  \rightarrow0
\]
and this justifies the approximation used in the paper.
\vspace{-0.8em}
\section{Proof of Proposition 1 }\label{proof1}
In order to prove Proposition 1, we consider the $i$-th entry
of the gradient of the objective function, given by
\[
\frac{\partial f_{0}}{\partial p_{i}}=\frac{\alpha_{i}+\alpha_{i}\sum_{j\neq
i}\beta_{j}p_{j}-\beta_{i}\sum_{j\neq i}\alpha_{j}p_{j}}{(1+\beta_{i}%
p_{i}+\sum_{j\neq i}\beta_{j}p_{j})^{2}}.
\]
When we treat the variables $\{p_{j}\}_{j\neq i}$ fixed, the $i$-th gradient
can be expressed as a function of $p_{i}$ in the form $\frac{\xi_{i}}%
{(\beta_{i}p_{i}+\zeta_{i})^{2}}$, where $\zeta_{i}>0$ and $\xi_{i}$ are
constants. Depending on the sign of $\xi_{i}$, the gradient is always negative
or positive, i.e. the function is monotonically decreasing or increasing in
each $p_{i}$ \footnote{For $\xi_{i}=0$, the function is constant in $p_{i}$,
then we let $p_{i}=0$.}. Since the objective function cannot be maximized at
$0<p_{i}<P_{i}$, the solution of (\ref{QuasiLP}) is achieved only at one of
the vertices. The second part follows directly from the monotonicity of the
function in each component $p_{i}$. Namely, the solution is achieved by the
vertex at which the objective function cannot further increase beyond the thresholds.

%%%%%%%%%%%%%%%%%%%%%%%%%%%%%%%%%%%%%%%%%%%%%%%%%%%%%%%%%%%%%%%%%%%%%%%%%%%%%%%%%%%%%%%%%%%%

\vspace{-0.8em}
\section{Proof of Corollary 1 : a closed-form solution of $M=2$ under perfect CSIT} \label{proof2}

From Proposition 1, we see immediately that the power allocation of two relays
depend on the sign of
\[
\xi_{1}(p_{2})=\alpha_{1}+\Delta p_{2},\;\;\xi_{2}(p_{1})=\alpha_{2}-\Delta
p_{1}%
\]
where we let $\Delta=\alpha_{1}\beta_{2}-\beta_{1}\alpha_{2}=|g_{1}g_{2}%
|^{2}(|h_{1}|^{2}-|h_{2}|^{2})$. Moreover, it is sufficient to check the sign
of $\xi_{1}$ and $\xi_{2}$ at each vertex to determine the optimum power
allocation. Table \ref{tab:gradient} summarizes the optimal solution and the
conditions ;  the optimal solution is given by
$(P_{1},0)$ if and only if $\xi_{2}(P_{1})=\alpha_{2}-\Delta P_{1}<0$
holds
while it is given by a vertex $(0,P_{2})$ if and only if we
have $\xi_{1}(P_{2})=\alpha_{1}+\Delta P_{2}<0$.
Finally, both relays are activated if $\frac{-\alpha_{1}}{P_{2}}<\Delta
<\frac{\alpha_{2}}{P_{1}}$.
These inequalities yield (\ref{2relayOnOff}).

%%%%%%%%%%%%%%%%%%%%%%%%%%%%%%%%%%%%%%%%%%%%%%%%%%%%%%%%%%%%%%%%%%%%%%%%%%%%%%%%%%%%%%%%%%%%

\vspace{-0.8em}
\section{Proof of Proposition 2 : convergence of on-off gradient algorithm}\label{proof3}

We have to first prove that the objective is non-decreasing, i.e.
$f(\pv^{(n+1)})\geq f(\pv^{(n)})$ for any iteration $n$. Identifying that the
update in (\ref{Update}) is nothing than a discrete steepest ascent algorithm
with a fixed step size, the objective always increases. It remains to prove
that the converged point is the global maximum. In other words, the stopping
criterion above is sufficient to guarantee a global convergence. It is not
difficult to see that there is a \textit{unique} $\pv^{\star}$ satisfying the
condition (\ref{SufficientCondition}) such that the signs of the gradients and
the powers match. Otherwise, we can always increase the objective by switching
on (off) the power with a positive (negative) gradient.

%%%%%%%%%%%%%%%%%%%%%%%%%%%%%%%%%%%%%%%%%%%%%%%%%%%%%%%%%%%%%%%%%%%%%%%%%%%%%%%%%%%%%%%%%%%%

\vspace{-0.8em}
\section{Proof of Proposition 3 : waterfilling solution under partial CSIT}\label{proof4}
Since $J(\widetilde{\pv})$ is a strictly concave function of
$\widetilde{\pv}$, we solve the KKT conditions which are necessary and
sufficient for optimality.
By letting $\lambda_{i}\geq0$ a Lagrangian variable associated to the individual power constraint $p_i\leq P_i$,
we obtain the KKT conditions given by
\begin{align}
\label{KKT}1- \frac{M\gamma_{gi}\exp(\tilde{p}_{i})}{1+ \sum_{j}\gamma
_{gj}\exp(\tilde{p}_{j})} =\lambda_{i},\;\; i =1,\dots,M
\end{align}
Summing the above equation over all $i$ and defining $I=\sum
_{j}\gamma_{gj}\exp(\tilde{p}_{j})$ and $\mu=\frac{1}{\sum_{i} \lambda_{i}}$
we obtain
\begin{equation}
\label{C}I= M\mu-1
\end{equation}
It follows that $\mu$ is lower bounded by $\mu^{\min}=\frac{1}{M}$ and upper
bounded by $\mu^{\max}=\frac{1+\sum_{j}\gamma_{gj}P_{j}}{M}$. Plugging
(\ref{C}) into (\ref{KKT})
and using the inequality $\lambda_{i}\geq0$, we readily obtain the optimal power given in (\ref{WFSolution}).

It remains to determine the optimal \textit{water level} $\tilde{\mu}^{\star}$ that maximizes the objective function $J$ (note that the individual power constraint is always satisfied for any $\mu$).
To this end, we define $\Cc(\tilde{\mu})$ and $\overline{\Cc}(\tilde
{\mu})$ as
\begin{align*}
\Cc(\tilde{\mu})  &  =\{i|\tilde{p}_{i}=\tilde{\mu}-\tilde{\gamma}_{gi}\},\;\;\;
\overline{\Cc}(\tilde{\mu}) =\{i|\widetilde{p_{i}}=\widetilde{P_{i}}\}
\end{align*}
Plugging (\ref{WFSolution}) into (\ref{LogObjective}), the function can be
expressed in terms of $\tilde{\mu}$
\begin{equation}\label{Jmu}
J(\tilde{\mu})=|\Cc(\tilde{\mu})|\tilde{\mu}+\sum_{i\in\overline{\Cc}%
(\tilde{\mu})}\tilde{P}_{i}-M\ln\left(  1+|\Cc(\tilde{\mu})|\exp(\tilde{\mu
})+\sum_{i\in\overline{\Cc}(\tilde{\mu})}\gamma_{gi}\exp(\tilde{P}%
_{i})\right)  +\sum_{i=1}^{M}\ln a_{i}%
\end{equation}
where $|\Cc|,|\overline{\Cc}|$ denotes the cardinality of the set
$\Cc,\overline{\Cc}$ respectively. Since the function $J(\tilde{\mu})$ is
strictly concave in $\tilde{\mu}$,
the optimal $\tilde{\mu}$ must satisfy $\frac{\partial J}{\partial\tilde{\mu}%
}=0$ where
\begin{equation}\label{DerivativeMu}
\frac{\partial J}{\partial\tilde{\mu}}=|\Cc(\tilde{\mu})|\left(  1-\frac
{M\exp(\tilde{\mu})}{1+|\Cc(\tilde{\mu})|\exp(\tilde{\mu})+\sum_{i\in
\overline{\Cc}(\tilde{\mu})}\gamma_{gi}\exp(\tilde{P}_{i})}\right)
\end{equation}
which yields
\begin{equation}\label{MuSolution}
\exp(\tilde{\mu})=\frac{1+\sum_{i\in\overline{\Cc}(\tilde{\mu})}\exp(\tilde
{P}_{i}+\tilde{\gamma}_{gi})}{|\overline{\Cc}(\tilde{\mu})|}%
\end{equation}
for $|\overline{\Cc}(\tilde{\mu})|>0$. Notice that for $|\overline{\Cc}%
(\tilde{\mu})|=0$, it can be shown that the objective function
is a monotonically increasing concave function and maximized at
$\tilde{\mu}^{\max}$. By
sorting $\{P_{i}\gamma_{gi}\}$ in an
increasing order according to the permutation (\ref{Permutation}), we remark that the RHS of (\ref{MuSolution}) has
at most $M$ possible values \footnote{Some of the $M$ values might be
unfeasible if they are not in the domain $\tilde{\mu}\in\lbrack\tilde{\mu
}_{\min},\tilde{\mu}_{\max}]$.} in (\ref{MuCandidate})
and we choose the optimal $\tilde{\mu}^{\star}$ according to (\ref{OptimalLevel}).

%%%%%%%%%%%%%%%%%%%%%%%%%%%%%%%%%%%%%%%%%%%%%%%%%%%%%%%%%%%%%%%%%%%%%

\vspace{-0.8em}
\section{Proof of Corollary 2 : a closed-form solution of $M=2$ under partial CSIT }\label{proof5}
Without loss of generality, we assume $P_{1}\gamma_{g,1}%
<P_{2}\gamma_{g,2}$. We recall that the possible values of the
water level (\ref{MuCandidate}) for $M=2$ are
\[
\mu_{1}=1+P_{1}\gamma_{g,1}\;\;\mu_{2}=\mu^{\mathrm{max}}=\frac{1+P_{1}%
\gamma_{g,1}+P_{2}\gamma_{g,2}}{2}.
\]
First we consider the case $\mu_{1}<\mu_{2}$. This inequality reduces to
$P_{2}\gamma_{g,2}>P_{1}\gamma_{g,1}+1$, and further yields %$\mu_{2}%
%<P_{2}\gamma_{g,2}$. Then we have
\[
P_{1}\gamma_{g,1}<\mu_{1}<\mu_{2}<P_{2}\gamma_{g,2}.%
\]
Obviously we obtain $p_{1}=P_{1}$ no matter which of the two values is the
optimum water level. It is not difficult to see that the optimal water level is given by $\mu_{1}$
by comparing the two objective values $J(\tilde{\mu_{1}})$ and $J(\tilde{\mu_{2}})$.
Hence the amplifier power of relay 2 is
$p_{2}=\frac{\mu_{1}}{\gamma_{g2}}=\frac{1+P_{1}\gamma_{g1}}{\gamma_{g2}}$.
Next we consider the case $\mu_{2}<\mu_{1}$. This inequality is equivalent to
$P_{2}\gamma_{g,2}<P_{1}\gamma_{g,1}+1$, and yields $P_{2}\gamma_{g,2}<\mu
_{2}$. In this case, the water level (either $\mu_{1}$ or $\mu_{2}$) is larger
than $P_{1}\gamma_{g,1}$ and $P_{2}\gamma_{g,2}$. Hence, the algorithm lets
both relays transmit at the maximum power, $p_{1}=P_{1}$ and $p_{2}=P_{2}$. In
summary, we have the following power allocation possibilities:

\begin{itemize}
\item $p_{1} =P_{1}$ and $p_{2}=\frac{1+P_{1} \gamma_{g1}}{\gamma_{g2}}<P_{2}$
if $P_{2} \gamma_{g2}>P_{1} \gamma_{g,1}+1$

\item $p_{1} =P_{1}$ and $p_{2}=P_{2}$ if $P_{2} \gamma_{g2}<P_{1}
\gamma_{g,1}+1$
\end{itemize}

By symmetry, when $P_{1} \gamma_{g1}> P_{2} \gamma_{g2}$ we obtain

\begin{itemize}
\item $p_{1}=P_{1}$ and $p_{2}=P_{2}$ if $P_{2}\gamma_{g2}<P_{1}\gamma
_{g,1}<P_{2}\gamma_{g2}+1$

\item $p_{1} =\frac{1+P_{2}\gamma_{g2}}{\gamma_{g1}}<P_{1}$ and $p_{2}=P_{2}$
if $P_{1}\gamma_{g1}>P_{2} \gamma_{g2}+1$
\end{itemize}

\begin{figure}[n]
\begin{center}
\epsfxsize=3in \epsffile{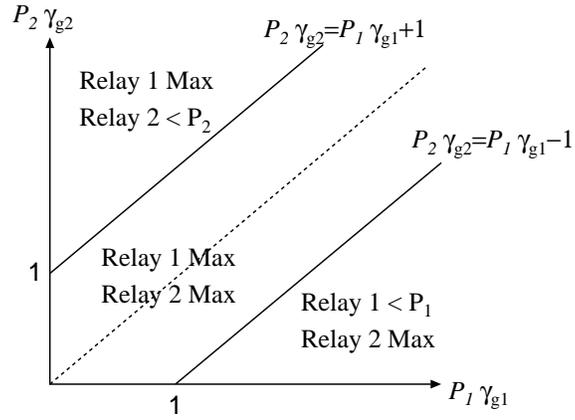}
\end{center}
\caption{Power allocation region as a function of $P_{i}\gamma_{gi}$}%
\label{fig:PRegion}%
\end{figure}

\begin{figure}[n]
\begin{center}
\epsfxsize=4in \epsffile{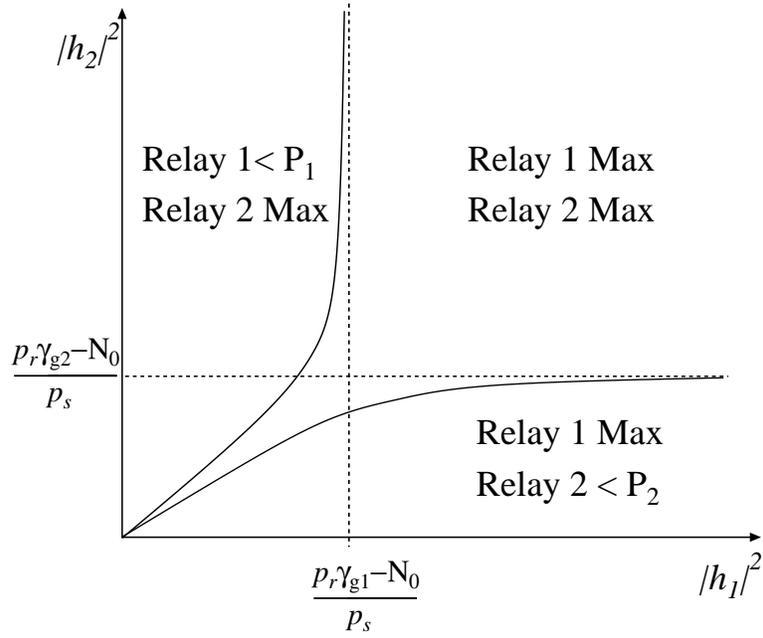}
\end{center}
\caption{Two-relay power region under partial CSIT}%
\label{fig:HRegion}%
\end{figure}

These conditions and the corresponding power allocation are summarized in
(\ref{3Subregion}) and depicted in Fig. \ref{fig:PRegion}. Since the parameter
$P_{i}\gamma_{gi}$ does not provide an easy interpretation (note that $P_{i}$
is a function of $\hv$), we express the power allocation region in terms of
$\gammav_{g}$, $\hv$ explicitly. The subregion where only relay 1, 2 is allocated
its maximum power is given respectively by
\[
|h_{2}|^{2}<\frac{p_{r}\gamma_{g2}\left(  |h_{1}|^{2}+\frac{N_{0}}{p_{s}%
}\right)  }{p_{s}|h_{1}|^{2}+p_{r}\gamma_{g1}+N_{0}}-\frac{N_{0}}{p_{s}},\;\; \;
|h_{1}|^{2}<\frac{p_{r}\gamma_{g1}\left(  |h_{2}|^{2}+\frac{N_{0}}{p_{s}%
}\right)  }{p_{s}|h_{2}|^{2}+p_{r}\gamma_{g2}+N_{0}}-\frac{N_{0}}{p_{s}}%
\]
These conditions yield the power allocation region in terms of $\hv$ in Fig.
\ref{fig:HRegion}.

%%%%%%%%%%%%%%%%%%%%%%%%%%%%%%%%%%%%%%%%%%%%%%%%%%%%%%%%%%%%%%%%%%%

\begin{figure}[n]
\begin{center}
\epsfxsize=3.5in \epsffile{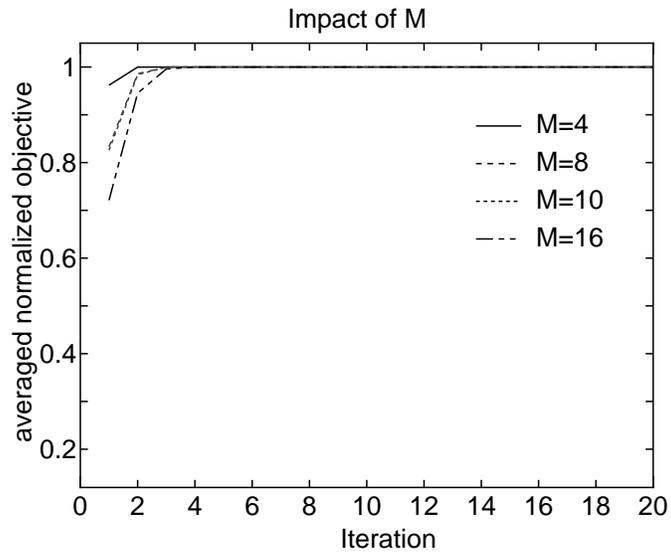}
\end{center}
\caption{Convergence of on-off algorithm for different $M$}%
\label{fig:ConvergenceM}%
\end{figure}

\begin{figure}[n]
\begin{center}
\epsfxsize=2.4in \epsffile{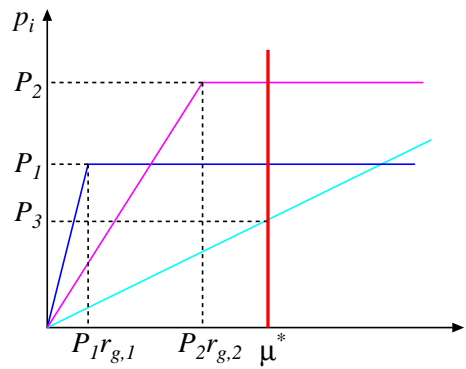}
\end{center}
\caption{Proposed waterfilling solution with $M=3$}%
\label{fig:KKT}%
\end{figure}

\begin{table}[n]
\centering
\begin{tabular}
[c]{c|cc|c}%
vertex & $\xi_{1}$ & $\xi_{2}$ & $\Delta$\\\hline
$(P_{1},0)$ & + & - & $(\alpha_{2}/P_{1},\infty]$\\\hline
$(P_{1},P_{2})$ & + & + & $(-\alpha_{1}/P_{2},\alpha_{2}/P_{1})$\\\hline
$(0,P_{2})$ & - & + & $[-\infty,-\alpha_{1}/P_{2})$\\\hline
\end{tabular}
\caption{Optimal solutions and corresponding conditions}%
\label{tab:gradient}%
\end{table}

\section*{Acknowledgment}

This work was partially supported by the Generalitat de Catalunya under grant
SGR2005-00690 and by the European Commission under project IST-6FP-033533
(COOPCOM).
\bibliographystyle{IEEEtran}
\bibliography{RelayDSTC-arxiv}

\end{document}